\documentclass[letterpaper]{article}
\usepackage[preprint]{aaai2027}
\usepackage[hyphens]{url}
\usepackage{graphicx}
\usepackage{natbib}
\usepackage{booktabs}
\title{Prediction Is Not Memory: Dual-Timescale Gated Profile Writing for Persistent User Modeling}
\author{
    Ziyide Li\\
}
\affiliations{
    School of Data Science and Intelligent Media\\
    Communication University of China\\
    \texttt{202313103009@cuc.edu.cn}
}

\begin{document}

\maketitle

\begin{abstract}
Persistent user profiles increasingly serve as reusable memory in recommender systems, but common update pipelines conflate two decisions: predicting an interaction and deciding whether it should persist in the profile. After an interaction is observed, many systems treat it as evidence for updating durable user state. This assumption can be harmful when the event reflects transient context, exploration, exposure, or short-term satisfaction rather than stable preference formation. We formulate this boundary as \emph{selective profile-write control}: after an observed interaction, a system should decide whether, and how strongly, to write it into the persistent profile. We introduce a chronological near/far offline protocol in which near-future evidence provides weak write-risk supervision and far-future evidence is reserved for evaluation. We instantiate the controller as SPW-Gate, a lightweight write-risk gate using long-term, short-term, and candidate-profile drift features. On MicroLens-100K, write-all updating hurts far-future profile alignment in 22.45\% of test cases under the main protocol. SPW-Gate reduces far hurt to about 14.5\% while preserving about 77\% write coverage. Matched-coverage controls and prediction-confidence baselines show that the gain is not merely a by-product of writing less, and that next-item confidence is not a sufficient proxy for persistent write validity.
\end{abstract}

\section{Introduction}

Modern recommender systems increasingly operate as stateful personalization pipelines rather than isolated next-item predictors. Established sequential and interest-evolution models update user representations from interaction histories and candidate-aware interests \cite{hidasi2016gru4rec,kang2018sasrec,sun2019bert4rec,zhou2018din,zhou2019dien}. Recent industrial and profile-oriented work extends this direction to long, heterogeneous, and non-stationary behavior streams, explicit user profiles, and longer-term user objectives \cite{zhai2024actions,si2024twinv2,feng2024context,han2025mtgr,zhou2024languageprofiles,gao2024langptune,liu2024dt4ier,wang2023streamingctr}. Together, these developments mark a shift from one-step recommendation toward systems that maintain user state over time.

This progress has made user modeling more expressive. Modern sequence models can attend to longer histories, profile-based methods can compress behavior into interpretable summaries, and long-term-objective models can move beyond immediate engagement. Yet many pipelines still share an implicit update assumption: once a user interacts with an item, the event becomes useful evidence for updating the user representation. That assumption is natural for short-term prediction, where fresh behavior can improve immediate relevance. It is less reliable for persistent profile writing, where the same update may shape later retrieval, ranking, explanation, and user understanding.

The missing boundary is therefore a pipeline-level decision boundary, not simply a need for another ranking architecture. Interaction prediction and profile persistence are distinct decisions, but update pipelines often couple them. Prediction asks whether an interaction will occur. Persistence asks whether that interaction should modify durable user memory. Work on implicit-feedback denoising and robust recommendation shows that observed interactions may contain false positives, hard samples, exposure effects, and other unreliable signals \cite{wang2021denoising,wang2022robust,song2024llmhd,liu2025crossdenoise}. As training examples, these events are noisy positives. Once committed to a profile store, they can distort the durable representation. A short-term response, exploratory action, trend-driven exposure, or momentary satisfaction may then be written as stable preference formation.

This distinction is consistent with recent work on AI long-term memory and brain-inspired learning, which revisits the separation between transient context and durable memory \cite{he2024longmemory,kabir2025timescale}. We do not claim to implement a cognitive architecture. Instead, the fast-response versus durable-memory distinction motivates a system boundary for recommendation: not every short-term response should be consolidated into persistent user memory.

We study this boundary as \emph{selective profile-write control}. After an interaction is observed, should it be written into the persistent profile, and if so, how strongly? This formulation imposes two requirements. First, the write decision must be learned without using the future evidence by which it will later be judged. We therefore use a chronological near/far protocol, where the near future provides weak supervision and the far future is reserved for evaluation. Second, the controller must identify cases where a candidate is supported by transient context but weakly supported by the durable profile. Its features should therefore expose long-term evidence, short-term evidence, and candidate-profile drift.

We propose \emph{Selective Profile-Write Gate} (SPW-Gate), a lightweight write-risk controller placed before the persistent profile update. SPW-Gate scores each candidate interaction using decision-time features from the long-term profile, the recent short-term profile, and their drift. Validation-selected thresholds convert the risk score into a write weight. We instantiate the scorer with logistic regression and a small MLP, denoted SPW-Gate-LR and SPW-Gate-MLP. These models are not recommendation backbones; they are lightweight implementations of the profile-write controller.

On MicroLens-100K \cite{ni2023microlens}, write-all updating can be harmful. Under the main \texttt{near1\_far1} protocol, it hurts far-future profile alignment in 22.45\% of test cases. SPW-Gate reduces this hurt to approximately 14.5\% while preserving approximately 77\% write coverage. Matched-coverage controls show that the improvement is not simply a consequence of writing fewer candidates. Prediction-confidence baselines derived from GRU, SASRec, and BERT4Rec \cite{hidasi2016gru4rec,kang2018sasrec,sun2019bert4rec} are near-random for identifying write risk, whereas direct write-risk gates provide a stronger signal. These results support the central claim that prediction confidence is not a sufficient substitute for persistent profile-write validity.

The contributions are therefore deliberately scoped:
\begin{itemize}
    \item We formulate post-interaction profile writing as a separate control problem from next-item prediction.
    \item We introduce a chronological near/far offline protocol for detecting harmful persistent writes without using the far future as model input.
    \item We show that a lightweight SPW-Gate reduces far-future profile hurt under substantial write coverage, with fixed-coverage controls against the explanation that it only writes less.
    \item We provide mismatch evidence that ranking confidence and persistent write validity are related but not interchangeable.
\end{itemize}

\section{Related Work}

\subsection{Sequential Recommendation and User State}

Sequential recommendation models infer a user's next action from historical interactions. Recurrent models such as GRU4Rec \cite{hidasi2016gru4rec}, self-attentive models such as SASRec \cite{kang2018sasrec}, and bidirectional sequence models such as BERT4Rec \cite{sun2019bert4rec} improve temporal user-state representation. Interest-network models such as DIN and DIEN \cite{zhou2018din,zhou2019dien} further emphasize target-aware or evolving interests. These models primarily optimize prediction or ranking: given a history and a candidate or item set, estimate future interaction likelihood.

Our problem is complementary. We do not ask whether a sequence model can predict a candidate interaction. We ask whether an observed candidate should be committed to a persistent profile. This post-interaction write decision can use prediction models as signals, but it is evaluated by the candidate's effect on later profile alignment. The experiments therefore include ranking-confidence baselines, while keeping the main method as a profile-write controller rather than a ranking model.

\subsection{Implicit Feedback Noise and Robust Recommendation}

Implicit feedback is noisy \cite{wang2021denoising,wang2022robust}. Prior work studies denoising implicit signals, robust training, and cross-model agreement for recommendation \cite{wang2021denoising,wang2022robust,song2024llmhd}. Such work typically aims to learn better recommenders from noisy observed interactions. Selective profile-write control focuses on a narrower system boundary: after an interaction is observed, should it be allowed to modify a durable user representation? This boundary matters because a persistent profile can affect many later components beyond the immediate ranking objective.

\subsection{Persistent Profiles as Memory}

Persistent profiles serve as system memory \cite{zhou2024languageprofiles,gao2024langptune}. They compress historical behavior into a reusable representation \cite{zhou2024languageprofiles,gao2024langptune}. Memory, however, has a different failure mode from prediction \cite{liu2024dt4ier}. A model may correctly predict a short-term action and still distort the durable profile if that action is not representative of future interests \cite{wang2021denoising,wang2022robust,liu2024dt4ier}. We operationalize this distinction through a near/far split: near-future labels are used to learn write risk, while far-future alignment is reserved for evaluating the profile update.

\section{Problem Formulation}

\subsection{Profile-Write Decision}

For each user $u$, let $H_u = (i_1,\ldots,i_t)$ be the historical interactions before an observed candidate item $c_u$. Let $N_u$ denote a near-future set and $F_u$ denote a later far-future set. The chronological order is:
\[
H_u \rightarrow c_u \rightarrow N_u \rightarrow F_u .
\]
The model observes $H_u$ and $c_u$ at write time. It may use labels derived from $N_u$ during training and validation. It must not use $F_u$ for model selection or decision-time features. The far future is used only for evaluation.

Each item has multimodal representations. For a set of items $S$, the profile representation $p(S)$ is the normalized mean item representation, averaged across available modalities. The base profile is $p(H_u)$. A write policy produces a candidate weight $w_u \in [0,1]$, yielding an updated profile:
\[
p'_u(w_u) =
\mathrm{norm}\left(
\frac{|H_u| \cdot \bar{e}(H_u) + w_u e(c_u)}
{|H_u| + w_u}
\right),
\]
where $\bar{e}(H_u)$ is the unnormalized mean history embedding and $e(c_u)$ is the candidate embedding. A binary write policy is a special case with $w_u \in \{0,1\}$.

\subsection{Alignment, Hurt, and Coverage}

Let $A(p,S)$ be the mean cosine alignment between profile $p$ and the profile representation of a future set $S$, averaged across modalities. The far-future update is harmful when:
\[
A(p'_u(w_u), F_u) < A(p(H_u), F_u).
\]
The \emph{far hurt rate} is the fraction of users for which this condition holds. The \emph{far alignment} is the average $A(p'_u(w_u),F_u)$. The \emph{write coverage} is the fraction of candidates assigned nonzero write weight. The \emph{mean weight} is the average $w_u$.

These metrics must be read together. A no-update policy has zero hurt by construction, but it also has zero write coverage and lower far alignment. Conversely, write-all has full coverage but can write harmful interactions. A useful policy should reduce far hurt while preserving enough coverage and alignment to remain a meaningful profile writer.

\subsection{Near-Future Write-Risk Label}

We define a near-future write-risk label from the near window:
\[
y_u^{near} =
\mathbf{1}\left[A(p'_u(1), N_u) < A(p(H_u), N_u)\right].
\]
This label indicates whether writing the candidate appears harmful with respect to the near future. SPW-Gate predicts this label from decision-time features. Final evaluation uses far-future alignment and far hurt. The protocol therefore tests whether near write-risk supervision transfers to later profile validity.

\section{Selective Profile-Write Gate}

\subsection{Decision-Time Features}

SPW-Gate uses a compact set of features available before the persistent write:
\begin{itemize}
    \item log history length, which captures how much evidence the current profile contains;
    \item candidate-long similarity, the similarity between the candidate and the full historical profile;
    \item candidate-short similarity, the similarity between the candidate and the recent short-term profile;
    \item long-short similarity, measuring drift between durable and recent interests;
    \item short-to-long similarity gap, measuring whether the candidate is more aligned with recent context than with long-term history.
\end{itemize}
These features encode a simple risk pattern: a candidate may be supported by recent context but weakly grounded in the durable profile.

\subsection{Risk Scoring and Write Policy}

Given feature vector $x_u$, SPW-Gate estimates a write-risk score $r_u = g(x_u)$. We instantiate $g$ with two lightweight scorers:
\begin{itemize}
    \item \textbf{SPW-Gate-LR}: logistic regression with standardized features and class-balanced training.
    \item \textbf{SPW-Gate-MLP}: a two-layer MLP with hidden sizes 32 and 16.
\end{itemize}
For each scorer, a validation search chooses a risk threshold $\tau$ and reduced write weight $\lambda$. The policy is:
\[
w_u =
\left\{
\begin{array}{ll}
\lambda, & r_u \ge \tau,\\
1, & r_u < \tau.
\end{array}
\right.
\]
In the main selected policies, $\lambda$ is often zero, so the controller behaves as a selective write/block gate. We retain the weighted formulation because the profile update rule also supports partial writes.

\subsection{Baselines}

We compare against four groups of baselines.
\textbf{Degenerate references} include no-update and write-all. They define the two extremes of the control space.
\textbf{Classic update baselines} include constant candidate weights, sliding windows, and time-decay updates. These test whether simple profile-maintenance heuristics are sufficient.
\textbf{Matched random policies} write a random subset at the same coverage or downweighting rate as learned gates. These directly test whether learned selection matters beyond writing less.
\textbf{Prediction-confidence baselines} use popularity, Markov transition scores, and candidate logits from GRU, SASRec, and BERT4Rec rankers. These test whether next-item confidence can substitute for write-risk estimation.

\section{Experiments}

\subsection{Dataset and Protocol}

We evaluate on MicroLens-100K \cite{ni2023microlens}, a content-driven micro-video recommendation dataset. The processed data contains 705,174 interactions, 98,129 users, and 17,228 items. The mean, median, 90th percentile, and maximum sequence lengths are 7.19, 6, 11, and 203, respectively. Each item has multimodal features. We compute profile alignment by averaging cosine alignment over text, image, and video item representations.

The main protocol is \texttt{near1\_far1}. It contains 58,194 eligible user samples after excluding users without the required future windows. The near-risk positive rate is 0.2144, the full-sample write-all far-hurt rate is 0.2200, and near/far label agreement is 0.7096. The protocol has no near/far overlap cases or time-order violations. We use three user-level split seeds: 20260620, 20260621, and 20260622. Policy selection uses validation near-future metrics; reported main results are test averages.

We also report robustness on \texttt{next\_one}, \texttt{near1\_far3}, and \texttt{near2\_far1}. The \texttt{next\_one} horizon is diagnostic because its near and far endpoints coincide. The main claim relies on the chronological \texttt{near1\_far1} protocol.

\subsection{Evaluation Questions}

The experiments are organized around five questions:
\begin{itemize}
    \item \textbf{Q1}: Does write-all profile updating cause measurable far-future hurt?
    \item \textbf{Q2}: Can SPW-Gate reduce far hurt while retaining meaningful write coverage and alignment?
    \item \textbf{Q3}: Is the result more than a side effect of writing fewer candidates?
    \item \textbf{Q4}: Can prediction confidence replace write-risk estimation?
    \item \textbf{Q5}: Does selective writing contradict independent far-future retrieval utility?
\end{itemize}

\subsection{Main Write-Control Results}

\begin{table*}[t]
\centering
\begin{tabular}{lccccc}
\toprule
Method & Far Hurt & Far Align. & Mean Weight & Coverage & Benefit Ret. \\
\midrule
No update & 0.0000 & 0.7448 & 0.0000 & 0.0000 & 0.0000 \\
Constant weight 0.25 & 0.1001 & 0.7491 & 0.2500 & 1.0000 & 0.3673 \\
Constant weight 0.50 & 0.1363 & 0.7518 & 0.5000 & 1.0000 & 1.0000 \\
Constant weight 0.75 & 0.1762 & 0.7532 & 0.7500 & 1.0000 & 1.0000 \\
Random matched & 0.1747 & 0.7517 & 0.7810 & 0.7810 & 0.7818 \\
Heuristic gate & 0.1905 & 0.7532 & 0.8961 & 0.8961 & 0.9099 \\
\textbf{SPW-Gate-LR} & 0.1471 & 0.7531 & 0.7784 & 0.7784 & 0.8140 \\
\textbf{SPW-Gate-MLP} & 0.1454 & 0.7530 & 0.7745 & 0.7745 & 0.8110 \\
Write all & 0.2245 & 0.7537 & 1.0000 & 1.0000 & 1.0000 \\
\bottomrule
\end{tabular}
\caption{Main write-control results on \texttt{near1\_far1}. Constant-weight rows write every candidate with the same fractional strength; thus their coverage is 1.0 and mean weight captures update strength. No-update is a degenerate reference, not a usable memory policy.}
\label{tab:main}
\end{table*}

Table~\ref{tab:main} reports the main \texttt{near1\_far1} write-control results. The table should not be read as a single-column leaderboard over far hurt. No-update has zero hurt only because it writes nothing, and it loses 0.0089 absolute far alignment relative to write-all. Constant weights 0.25 and 0.50 also reduce hurt by uniformly weakening every candidate update; they are useful attenuation baselines, but they do not decide which interactions should enter persistent memory.

The main comparison is therefore among non-degenerate policies with comparable memory strength. Write-all reaches the highest far alignment, 0.7537, but also produces 22.45\% far hurt. SPW-Gate-LR and SPW-Gate-MLP reduce far hurt to 14.71\% and 14.54\%, while retaining roughly 78\% and 77\% write coverage. At similar write strength, matched random writing has 17.47\% far hurt and the constant-0.75 baseline has 17.62\% far hurt. SPW-Gate therefore improves the risk-utility tradeoff by selecting safer writes, not merely by applying a global attenuation factor.

\begin{figure}[tbp]
\centering
\includegraphics[width=0.95\linewidth]{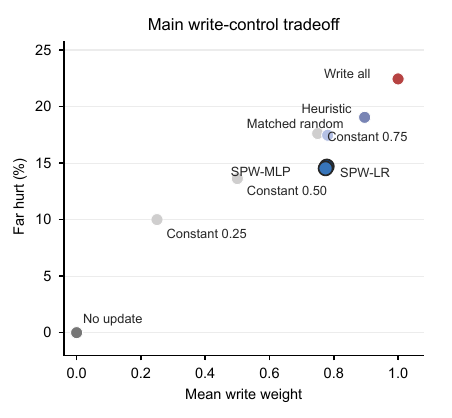}
\caption{Main risk-utility tradeoff on \texttt{near1\_far1}. Far hurt is plotted against mean write weight; error bars denote standard deviation across the three user-level split seeds.}
\label{fig:tradeoff_main}
\end{figure}

Figure~\ref{fig:tradeoff_main} visualizes the same comparison. The useful region is the middle of the plot: write-all sits at full update strength but high hurt, no-update sits at zero hurt but zero memory writing, and SPW-Gate occupies a selective operating point with much lower hurt and near write-all alignment.

\subsection{Classic Profile-Update Baselines}

Classic profile-maintenance rules provide additional context. The constant-weight baselines in Table~\ref{tab:main} and Figure~\ref{fig:tradeoff_main} show that uniform weakening can reduce hurt, but this comes with a different interpretation from selective writing. Weight 0.25 reaches 10.01\% hurt, yet it keeps only a quarter-strength candidate update and far alignment drops to 0.7491. Weight 0.50 reaches 13.63\% hurt, but still applies the same half-strength update to every interaction rather than identifying which interactions should persist. Weight 0.75 is closer to SPW-Gate's mean weight and has 17.62\% hurt, worse than both SPW-Gate variants.

Sliding-window and time-decay updates are less effective in this protocol. Last-5 sliding window reaches 33.44\% hurt; 30-day time decay reaches 44.95\%; last-3 sliding window reaches 49.86\%; and 7-day time decay reaches 73.76\%. These results suggest that recency alone is not a sufficient memory rule. The central question is not only how recent the memory should be, but whether the candidate interaction is safe to write into the persistent profile at all.

\subsection{Non-Degeneracy and Fixed-Coverage Pareto}

The main table alone does not establish non-degeneracy, because any conservative policy can lower hurt by writing less. We therefore run fixed-coverage controls. At each target coverage, the learned gates and random controls write the same fraction of candidates, after which we recompute exact far-future profile alignment.

Table~\ref{tab:coverage} reports far hurt at 70\%, 80\%, and 90\% target coverage. SPW-Gate-LR and SPW-Gate-MLP are consistently below random writing at the same coverage, and also below the heuristic gate. This addresses the main alternative explanation: the learned gate is not only suppressing writes; it ranks candidate writes by risk well enough to select safer interactions under the same write budget.

\begin{table}[t]
\centering
\begin{tabular}{lccc}
\toprule
Method & 70\% Cov. & 80\% Cov. & 90\% Cov. \\
\midrule
Random & 0.1604 & 0.1786 & 0.2016 \\
Heuristic & 0.1382 & 0.1642 & 0.1929 \\
\textbf{SPW-LR} & \textbf{0.1273} & \textbf{0.1526} & \textbf{0.1817} \\
\textbf{SPW-MLP} & \textbf{0.1279} & \textbf{0.1532} & \textbf{0.1832} \\
\bottomrule
\end{tabular}
\caption{Far hurt at fixed target coverage on \texttt{near1\_far1}. Learned gates outperform random writing under the same write budget.}
\label{tab:coverage}
\end{table}

\begin{figure}[t]
\centering
\includegraphics[width=0.95\linewidth]{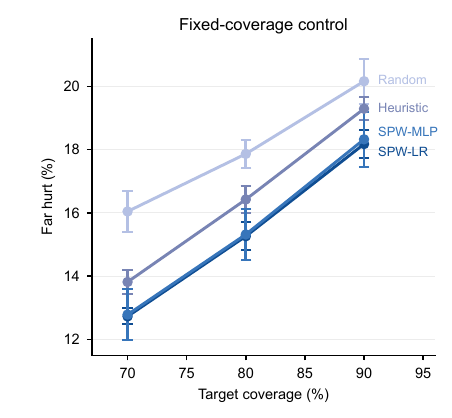}
\caption{Fixed-coverage control on \texttt{near1\_far1}. Error bars denote standard deviation across the three user-level split seeds.}
\label{fig:tradeoff_fixed_coverage}
\end{figure}

Figure~\ref{fig:tradeoff_fixed_coverage} shows the corresponding fixed-coverage curves. As the target coverage increases, far hurt rises for all methods, but the learned gates remain below random and heuristic controls across the evaluated coverage range.

\subsection{Prediction-Persistence Mismatch}

We next test whether prediction confidence can replace write-risk estimation. Table~\ref{tab:mismatch} reports two diagnostics. Risk AUC measures how well a score orders candidates by the near-future write-risk label. Top-decile far hurt measures the actual far-future hurt rate among the 10\% highest-scored candidates. A useful write-risk score should exceed random AUC and concentrate harmful writes in its high-risk decile.

Direct write-risk gates reach AUC around 0.62. Their top-risk deciles have far hurt around 0.39--0.41, compared with the overall write-all far-hurt rate of 0.2245. In contrast, Markov and popularity confidence are weak, and candidate ranking logits from GRU, SASRec, and BERT4Rec rankers remain close to random AUC. This supports the paper's core distinction: a score that helps predict the next item is not automatically a score that tells the profile store whether the interaction should become memory.

\begin{table*}[t]
\centering
\begin{tabular}{llccc}
\toprule
Source & Score & Risk AUC & Top-Decile Far Hurt & Overall Far Hurt \\
\midrule
SPW risk gate & MLP write-risk & 0.6206 & 0.4031 & 0.2245 \\
SPW risk gate & LR write-risk & 0.6198 & 0.4134 & 0.2245 \\
Neural write-risk gate & SASRec write-risk & 0.6204 & 0.3853 & 0.2245 \\
Neural write-risk gate & BERT4Rec write-risk & 0.6200 & 0.4037 & 0.2245 \\
Neural write-risk gate & GRU state-aug. write-risk & 0.6027 & 0.3755 & 0.2245 \\
Basic confidence & Markov probability & 0.5183 & 0.2795 & 0.2245 \\
Basic confidence & Popularity probability & 0.5140 & 0.2263 & 0.2245 \\
Ranking confidence & GRU candidate logit & 0.5090 & 0.2524 & 0.2245 \\
Ranking confidence & SASRec candidate logit & 0.5084 & 0.2386 & 0.2245 \\
Ranking confidence & BERT4Rec candidate logit & 0.5044 & 0.2369 & 0.2245 \\
\bottomrule
\end{tabular}
\caption{Prediction-persistence mismatch on \texttt{near1\_far1}. Candidate logits from next-item rankers are weak write-risk signals, while direct write-risk gates better identify interactions that later hurt persistent profile alignment.}
\label{tab:mismatch}
\end{table*}

\begin{figure*}[t]
\centering
\includegraphics[width=0.88\textwidth]{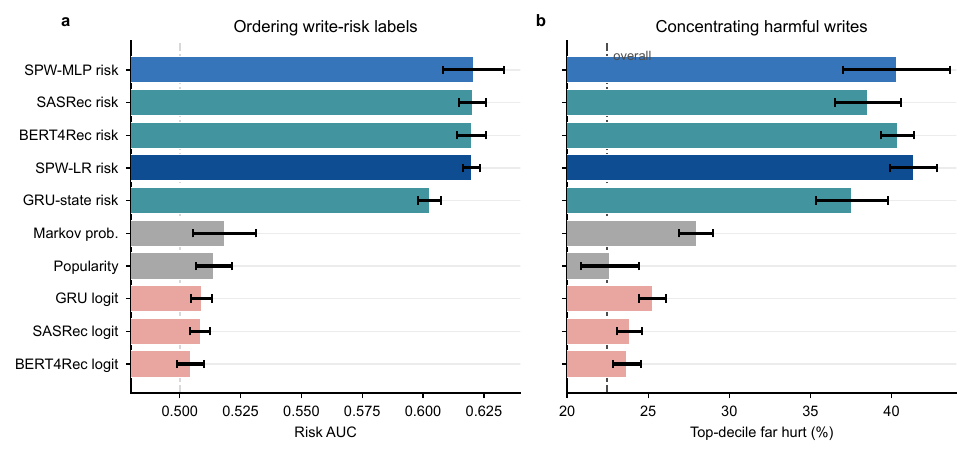}
\caption{Prediction-persistence mismatch on \texttt{near1\_far1}. (a) Risk AUC for near-future write-risk labels. (b) Far hurt in the top risk decile. The dashed reference in (b) marks the overall write-all far-hurt rate.}
\label{fig:mismatch}
\end{figure*}

Figure~\ref{fig:mismatch} visualizes the same pattern. Direct write-risk scores are above the random-AUC reference and concentrate far hurt in the top decile, whereas ranking-confidence scores stay near chance and only weakly separate harmful writes.

\subsection{Sequence Gates as Task-Adapted Comparisons}

The neural write-risk rows in Table~\ref{tab:mismatch} use GRU, SASRec, and BERT4Rec-style sequence encoders trained directly for the write-risk label. They should be separated from the ranking-confidence rows, which reuse candidate logits from next-item rankers. Under selected write-control policies on \texttt{near1\_far1}, the SASRec write-risk gate reaches 15.03\% far hurt, BERT4Rec reaches 15.27\%, GRU with state augmentation reaches 15.88\%, and the GRU sequence gate reaches 16.84\%. All reduce hurt relative to write-all.

This comparison clarifies the role of sequence models. When the same model family is trained for next-item ranking, its candidate logit is a weak write-risk signal. When trained directly for write risk, a sequence encoder can become a valid write controller. The lightweight SPW-Gate variants remain competitive with these task-adapted sequence gates, so the main effect in this offline proxy does not require a heavy sequence backbone.

\subsection{Horizon Robustness}

Table~\ref{tab:horizon} reports far hurt across additional horizons. The learned gates remain below write-all across all reported settings. The effect is strongest and most interpretable in the chronological \texttt{near1\_far1} protocol, which is the main setting. The \texttt{next\_one} horizon is diagnostic because its near and far endpoints coincide; it does not test temporal transfer.

\begin{table}[t]
\centering
\begin{tabular}{lcccc}
\toprule
Horizon & All & LR & MLP & Heur. \\
\midrule
\texttt{next} & 0.2062 & 0.1373 & 0.1464 & 0.1644 \\
\texttt{n1-f1} & 0.2245 & 0.1471 & 0.1454 & 0.1905 \\
\texttt{n1-f3} & 0.1409 & 0.0752 & 0.0899 & 0.1062 \\
\texttt{n2-f1} & 0.2314 & 0.1455 & 0.1485 & 0.1879 \\
\bottomrule
\end{tabular}
\caption{Far hurt across horizons. The main claim uses \texttt{near1\_far1}; other horizons are robustness and diagnostic settings.}
\label{tab:horizon}
\end{table}

\begin{figure}[!t]
\centering
\includegraphics[width=0.95\linewidth]{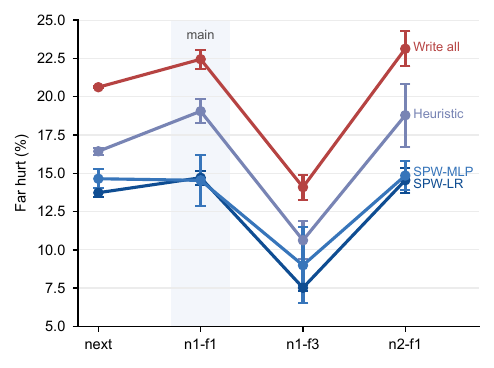}
\caption{Far hurt across horizons. The shaded region marks the main \texttt{near1\_far1} protocol. Error bars denote standard deviation across three user-level split seeds.}
\label{fig:horizon}
\end{figure}

Figure~\ref{fig:horizon} is the visual counterpart of Table~\ref{tab:horizon}. The shaded region marks the main protocol, and the error bars show variation across the three user-level split seeds.

\setlength{\floatsep}{6pt plus 1pt minus 1pt}
\setlength{\textfloatsep}{8pt plus 1pt minus 2pt}

\subsection{Full-Catalog Retrieval Compatibility}

We also evaluate whether selective writing is compatible with far-future retrieval over the full MicroLens item catalog. The updated profile is used as a query; far-future items are positives; history items and the observed candidate are masked. This experiment is a compatibility check, not evidence that SPW-Gate improves a production ranking stack.

\begin{figure*}[t]
\centering
\includegraphics[width=0.82\textwidth]{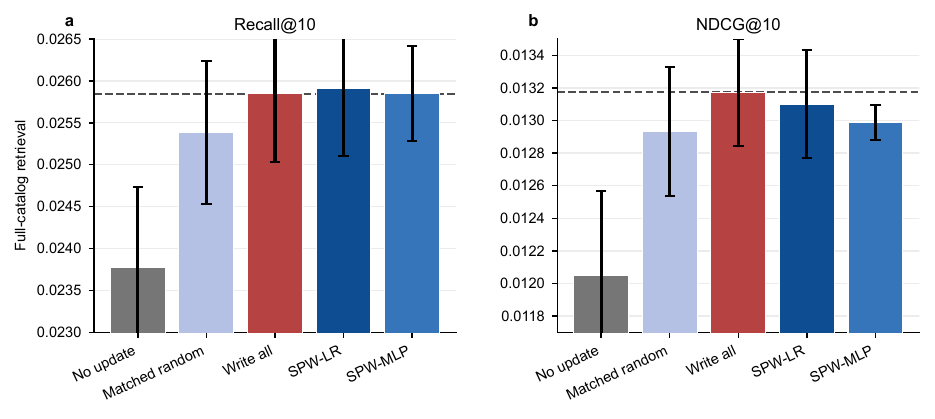}
\caption{Full-catalog far-future retrieval compatibility check on \texttt{near1\_far1}. Panels report Recall@10 and NDCG@10; dashed lines mark the write-all reference. Error bars denote standard deviation across three user-level split seeds.}
\label{fig:retrieval}
\end{figure*}

Retrieval metrics are close across non-degenerate write policies. Recall@10 is 0.02585 for write-all, 0.02591 for SPW-Gate-LR, 0.02585 for SPW-Gate-MLP, and 0.02539 for matched random. No-update is lower at 0.02378. NDCG@10 follows a similar pattern: 0.01317 for write-all, 0.01310 for SPW-Gate-LR, 0.01299 for SPW-Gate-MLP, and 0.01293 for matched random. Figure~\ref{fig:retrieval} shows these two metrics with seed-level uncertainty. Read conservatively, this supports compatibility with the independent retrieval check, not a large retrieval gain.

\subsection{Statistical and Calibration Checks}

Finally, we audit the stability and calibration of the selected write-risk gates. Table~\ref{tab:stats_calibration} summarizes two supporting diagnostics on \texttt{near1\_far1}. The first group reports user-clustered bootstrap differences against write-all over 1000 resamples. Negative hurt differences indicate lower far-future hurt. The second group reports Platt-calibrated risk-score quality on the test split.

\begin{table}[!t]
\centering
{\small
\setlength{\tabcolsep}{3pt}
\begin{tabular}{lcc}
\toprule
Diagnostic & SPW-Gate-LR & SPW-Gate-MLP \\
\midrule
Hurt diff vs write-all & -0.0774 & -0.0792 \\
95\% CI & [-0.0843, -0.0708] & [-0.0861, -0.0720] \\
Align. diff vs write-all & -0.0006 & -0.0007 \\
\midrule
Brier & 0.1621 & 0.1617 \\
Log loss & 0.5022 & 0.5010 \\
ECE@10 & 0.0157 & 0.0089 \\
ROC-AUC & 0.6198 & 0.6206 \\
Average precision & 0.2990 & 0.2965 \\
\bottomrule
\end{tabular}
}
\caption{Stability and calibration checks on \texttt{near1\_far1}. Hurt and alignment differences use a user-clustered bootstrap against write-all; calibration metrics use Platt-calibrated write-risk scores.}
\label{tab:stats_calibration}
\end{table}

The bootstrap intervals support the claim that hurt reduction is stable under user-level resampling: SPW-Gate-MLP reduces far hurt by 0.0792 on average, with a 95\% confidence interval from -0.0861 to -0.0720; SPW-Gate-LR reduces far hurt by 0.0774, with an interval from -0.0843 to -0.0708. The corresponding far-alignment differences are small and negative, -0.0007 for SPW-Gate-MLP and -0.0006 for SPW-Gate-LR, indicating that alignment is approximately maintained rather than improved.

The calibration audit is supporting rather than central. Since the main method uses risk scores for thresholded write control, the primary claim does not require perfectly calibrated probabilities. Still, after validation-fitted Platt calibration, both gates obtain similar Brier score, log loss, ROC-AUC, and average precision, with SPW-Gate-MLP showing the lower ECE@10. This suggests that the write-risk scores can be made reasonably interpretable when calibrated, while the main contribution remains selective write control.

\section{Limitations and Claim Scope}

This paper makes a bounded claim. Under a MicroLens offline near/far proxy, a lightweight selective profile-write controller reduces far-future profile hurt while preserving substantial write coverage. The results do not establish a production recommender improvement, claim state-of-the-art ranking, or prove a general theory of long-term human preference.

There are three important limitations. First, the current evidence is from one dataset family. Future work should add at least one cross-domain dataset and one exposure-aware sequential dataset. Second, the present ranking-confidence mismatch uses candidate logits from existing ranker artifacts, not full-catalog confidence features such as rank percentile, top-K membership, score entropy, or margin. Third, the current profile update is a one-step candidate write. Multi-step persistence, repeated writes, and explicit write/defer/block actions remain future work.

Within this scope, the evidence supports a narrower research story: prediction and memory writing should be separated, and a small write-risk gate can act as a practical control layer for persistent user profiles.

\section{Conclusion}

This paper separates two decisions that are often coupled in recommender-system update pipelines: predicting an interaction and deciding whether that interaction should become persistent profile memory. We formulated the latter as selective profile-write control, introduced a near/far offline protocol for evaluating harmful persistent writes, and instantiated the controller with a lightweight SPW-Gate. Across the main MicroLens protocol, write-all updating produces measurable far-future profile hurt, while SPW-Gate reduces hurt under substantial write coverage. Fixed-coverage controls show that this reduction is not only a consequence of writing fewer candidates. Prediction-persistence mismatch results further show that next-item confidence is a weak substitute for direct write-risk estimation. The claim is deliberately bounded: the results establish a credible offline proxy for one-step profile-write control, not a production ranking improvement or a complete theory of long-term preference. Within this scope, the evidence supports a practical design principle: persistent user memory should have its own write boundary.

\bibliography{spw_refs}

\end{document}